\documentclass[conference]{IEEEtran}
\IEEEoverridecommandlockouts
\usepackage{times}
\usepackage{url}
\usepackage{latexsym}

\usepackage{graphicx}
\usepackage{flushend}
\usepackage{algorithm}
\usepackage{algorithmic}
\usepackage{subfigure}

\usepackage{booktabs}
\usepackage{multirow}
\usepackage{xcolor}
\usepackage{amsmath}
\usepackage{tabularx}
\usepackage{enumerate}
\usepackage{amssymb}
\usepackage{amsmath}
\usepackage{amsthm}
\theoremstyle{definition}
\newtheorem{definition}{Definition}%[section]

\newcommand{\rarrow}[1]{\vec{#1\vphantom{b}}}

\begin{document}

\title{Pseudo-Implicit Feedback for Alleviating Data Sparsity in Top-K Recommendation \\}
%\author{Anonymous}

\author{
\IEEEauthorblockN{Yun He,  Haochen Chen\IEEEauthorrefmark{2}, Ziwei Zhu, James Caverlee} 
  
\IEEEauthorblockA{Department of Computer Science and Engineering, Texas A\&M University} 
\IEEEauthorblockA{\IEEEauthorrefmark{2}Department of Computer Science, Stony Brook University\\
} 
\{yunhe,zhuziwei,caverlee\}@tamu.edu; \IEEEauthorrefmark{2}haocchen@cs.stonybrook.edu
}
\maketitle

\begin{abstract}
% \ourmethod
%Top-k recommendation aims to identify the most preferred items for a user. 

We propose PsiRec, a novel user preference propagation recommender that incorporates pseudo-implicit feedback for enriching the original sparse implicit feedback dataset. Three of the unique characteristics of PsiRec are: (i) it views user-item interactions as a bipartite graph and models pseudo-implicit feedback from this perspective; (ii) its random walks-based approach extracts graph structure information from this bipartite graph, toward estimating pseudo-implicit feedback; and (iii) it adopts a Skip-gram inspired measure of confidence in pseudo-implicit feedback that captures the pointwise mutual information between users and items.
This pseudo-implicit feedback is ultimately incorporated into a new latent factor model to estimate user preference in cases of extreme sparsity. PsiRec results in improvements of 21.5\% and 22.7\% in terms of Precision@10 and Recall@10 over state-of-the-art Collaborative Denoising Auto-Encoders.  Our implementation is available at \url{https://github.com/heyunh2015/PsiRecICDM2018}. %Experiments conducted on three public datasets show that PsiRec outperforms traditional matrix factorization approaches as well as more sophisticated neural network approaches, effectively alleviating the data sparsity problem. For instance,

\end{abstract}


\section{Introduction}
%Recommender systems are critical for helping users select items from vast collections. For example, Amazon's recommenders help consumers make sense of a repository of more than 500 million products\footnote{https://www.scrapehero.com/many-products-amazon-sell-january-2018/}. Similar scenarios arise in digital media (e.g., movies, music), social networks, news aggregators, and job platforms. With the dearth of explicit ratings in many platforms -- e.g., only 15\% of consumers on Amazon write reviews\footnote{https://www.usatoday.com/story/tech/news/2017/03/20/review-you-wrote-amazon-priceless/99332602/} -- there is a push to exploit implicit feedback based on user-item interactions like clicks, purchases, and add-to-cart actions \cite{hu2008collaborative}. However, even such implicit feedback sources can suffer from sparsity, leading to poor performing recommenders. For example, in a sample of Amazon products described in Section~\ref{data sets} we find that most users have interacted with fewer than 7 out of 10,000+ items, meaning that models of user preference are subject to being underfit and error-prone.

%Hence, this paper focuses on the problem of top-K recommendation under conditions of extreme sparsity in implicit feedback datasets. 
We focus on the problem of top-K recommendation under conditions of extreme sparsity in implicit feedback datasets. The goal of top-K item recommendation is to generate a list of $K$ items for each user, typically by modeling the hidden preferences of users toward items based on the set of actual user-item interactions. While implicit feedback (like clicks, purchases) is often more readily available than explicit feedback, in many cases it too suffers from sparsity since users may only interact with a small subset of all items, leading to poor quality recommendation. %Hence, we aim to overcome such sparsity by carefully modeling pseudo-implicit feedback from the larger space of indirect, transitive user-item pairs. 

%In contrast to the cold start problem in which a new user or item has just entered the system, interaction sparsity here refers to the \textit{neighbor transitivity} \cite{su2009survey} problem that the interaction dataset is so sparse that users with similar preference may not be identified because they give no feedback on any of the same items. As a consequence, there is not enough history behavior of similar users to support recommendation.

Typically, there are two main strategies to alleviate the interaction sparsity problem. The first is to adopt a latent factor model to map users and items into the same low-rank dimensional space \cite{hu2008collaborative,koren2009matrix,cdae,rendle2009bpr,pan2008one,ncf}.  The second is to propagate user preferences using random walks to ``nearby'' items so that direct user-item pairs can propagate to transitive user-item pairs \cite{huang2004applying,itemrank,yildirim2008random,trustwalker,christoffel2015blockbusters}. In this paper, we propose a novel approach that combines the advantages of both latent factor and random walk models. 

%In this way, two users who may have given no feedback on the same item may share some similarities in the latent factor space.

Our key intuition is to carefully model \textit{pseudo-implicit feedback} from the larger space of indirect, transitive user-item pairs and embed this feedback in a latent factor model. Concretely, we propose PsiRec, a novel user preference propagation recommender that incorporates pseudo-implicit feedback for enriching the original sparse implicit feedback dataset. This pseudo-implicit feedback can naturally be estimated through an embedding model which we show explicitly factorizes a matrix where each entry is the pointwise mutual information value of a user and item in our sample of pseudo user-item pairs. We show how this pseudo-implicit feedback corresponds to random walks on the user-item graph. To model  confidence in pseudo-implicit feedback, we adopt a Skip-gram \cite{mikolov2013distributed} inspired measure that captures the pointwise mutual information between users and items.
 
 %Specifically, the pseudo-implicit feedback is achieved by generating random walks on the bipartite graph between users and items. 
%Evaluations are conducted on three very sparse and publicly accessible datasets from two large online shopping platforms: Amazon and Tmall. 
Experiments over sparse Amazon and Tmall datasets show that our approach outperforms matrix factorization approaches as well as more sophisticated neural network approaches and effectively alleviates the data sparsity problem. Further, we observe that the sparser the datasets are, the larger improvement can be obtained by capturing the pseudo-implicit feedback from indirect transitive relationships among users and items.%\textbf{Put some findings here}

\section{Related Work}
\label{related work}
There is a rich line of research on top-K recommendation for implicit feedback datasets
\cite{hu2008collaborative,cdae,rendle2009bpr,pan2008one,ncf}, where typically positive examples
are extremely sparse.
One family of approaches enriches the interactions between users and items via exploring the transitive, indirect associations that are not observed in the training data \cite{huang2004applying,itemrank,yildirim2008random,trustwalker,christoffel2015blockbusters}.
Huang et al. \cite{huang2004applying} compute the association between a user and an item as the sum of the weights of all paths connecting them.
%Papagelis et al. \cite{papagelis2005alleviating} explore transitive similarities between users using trust inferences.
ItemRank \cite{itemrank} employs PageRank on the correlation graph between items to
infer the preference of a user.
$P^3$ \cite{cooper2014random} ranks items according to the third power of the transition matrix.%of $G$.

% The most important difference between our method and the above approaches is that
% we do not directly compute the ranking of items based on the association
% between users and items.
% Instead, we exploit such association with latent factor (LF) models.
As another line of research, latent factor (LF) models learn to assign each user a vector of latent user factors
and each item with a vector of latent item factors.
These dense latent factors enable similarity computations between arbitrary pairs of users and items, and thus can address the data sparsity problem.
As a representative type of LF models, matrix factorization (MF) methods
model the preference of user $u$ over item $i$ as the inner product of their latent representations \cite{hu2008collaborative,koren2009matrix}.
In terms of optimization, they aim to minimize the squared loss between the predicted preferences and the ground-truth preferences.
A number of works can be seen as a variation of MF methods by employing different models to learn
the user-item interaction function \cite{cdae,ncf} or using different loss functions \cite{rendle2009bpr,rennie2005fast}.
Our method brings together these two lines of research by exploiting both direct and indirect associations between users and items with a matrix factorization model.

\section{Preliminaries}
\label{preliminaries}
%\subsection{Preliminaries}
%We focus on the problem of top-K recommendation under conditions of extreme sparsity in implicit feedback datasets. Implicit feedback is the observation of the users' action on items, like clicks, add-to-cart actions, and purchases. Unlike explicit feedback, like review ratings, which clearly express the user preference, implicit feedback indirectly reflects the user preference. For example, the observation that a user buys an item implies that the user probably likes the item. Because high quality explicit feedback is not always available, recommendations based on implicit feedback receive more and more attention recently \cite{hu2008collaborative}.

%\subsection{Top-K Item Recommendation}
Let $U$ and $I$ denote the set of users and items respectively. Additionally, let $M$ and $N$ represent the number of users and items. We also reserve $u$ to represent a user and $i$ to represent an item. We define the user-item interaction matrix as $\textbf{R} \in \mathbb{R}^{M \times N}$, and denote the implicit feedback from $u$ on $i$ as $r_{ui}$. In many cases, implicit feedback datasets only contain the logs of binary interactions where $r_{ui}=1$ indicates $u$ purchased $i$ and $r_{ui}=0$ the otherwise. Therefore, we focus on binary interactions in this paper. Non-binary interactions can be handled with slight modifications to our approach. Based on these interactions, top-K item recommendation generates a list of $K$ items for each user, where each item $i$ is ranked by the (hidden) preference of $u$, denoted as $p_{ui}$. The preference matrix for all users and items is denoted as $\textbf{P} \in \mathbb{R}^{M \times N}$.

One traditional approach \cite{koren2009matrix} for estimating this user preference is through latent factor models like matrix factorization.  We define $\textbf{x}_{u}$ to be the latent factor vector for $u$ and $\textbf{y}_{i}$ to be the latent factor vector for $i$. Matrices $\textbf{X} \in  \mathbb{R}^{M \times K}$ and $\textbf{Y} \in \mathbb{R}^{N \times K}$ represent the latent factor matrix for users and items accordingly. User preferences can be estimated as the following inner product: $\hat{p}_{ui}=\textbf{x}_{u}^{T}\textbf{y}_{i} $. Model parameters can then be learned by minimizing the squared error between $r_{ui}$ and $\hat{p}_{ui}$: $\ell = \sum_{u,i}(r_{ui}-\hat{p}_{ui})^2$. However, in case of extremely sparse user-item interactions, even such latent factor models may face great challenges: the number of direct user-item pairs ($r_{ui}=1$) may be too few to learn meaningful latent factors. %For example, in an Amazon dataset of purchases in the Cell Phone category, there are 27,879 users and 10,429 items (see Table~\ref{data statistics}). On average, each user has purchased only 6.97 items out of the entire 10,429 items; further, more than 88\% of all users have purchased fewer than 10 items. How to uncover the user preferences for the remaining 10,000+ items can be challenging. Indeed, through experiments presented in Section~\ref{Robustness of Our Approach on Very Sparse Dataset}, we find that traditional matrix factorization approaches as well as more sophisticated neural network approaches  suffer over such sparse datasets.

%It will be shown that the sparser the datasets are, the worse performance is achieved by collaborative filtering methods. For example, for Amazon Toys and Games dataset, when density of $\textbf{R}$ drops from 0.072\% to 0.017\%, the performance of matrix factorization drops from 0.02532 to 0.00099 in terms of recall@10. 

%\subsection{The Challenge of Extreme Sparsity}
%Taking Amazon Cell Phones Dataset as the example in Table \ref{data statistics}, on average, each user purchased only 6.97 items out of the whole 10,429 items and 88.5\% of users have fewer than 10 purchase. In this case, users have feedback on so small portion of items that the users' preference cannot be learned sufficiently. This argument is verified in Section \ref{Robustness of Our Approach on Very Sparse Dataset}. It will be shown that the sparser the datasets are, the worse performance is achieved by collaborative filtering methods. For example, for Amazon Toys and Games dataset, when density of $\textbf{R}$ drops from 0.072\% to 0.017\%, the performance of matrix factorization drops from 0.02532 to 0.00099 in terms of recall@10. 

\section{PsiRec: Pseudo-Implicit Feedback}
\label{PsiRec paragraph}
In this section, we propose a general approach for enhancing top-K recommenders with \textit{pseudo-implicit feedback}. Our key intuition is that the original sparse interaction matrix $\textbf{R}$ can be enriched by carefully modeling transitive relationships to better estimate user preference. Concretely, our approach can be divided into steps given as follows:
\begin{itemize}
\item First, we view the user-item interactions as a bipartite graph and propose to model pseudo-implicit feedback from this perspective.
\item Second, we propose a random walks-based approach to extract evidence of user-item closeness from this bipartite graph toward estimating pseudo-implicit feedback.
\item Third, we describe two strategies to measure the confidence of pseudo-implicit feedback based on these sampled user-item pairs from the bipartite graph.
\item Finally, we develop a latent factor model to estimate user preference from this pseudo-implicit feedback.
\end{itemize}

%the transitive relationships is better to estimate the user preference.

%In this section, we propose to overcome extreme sparsity in top-k item recommendation through exploiting transitive relationships among users and items. 

%We first provide an overview of our approach. Then, a random walks-based approach is presented to sample the indirect user-item pairs. After that, a neural embedding method is proposed to measure the pseudo-implicit feedback. Finally, a latent factors model is proposed to estimate the user preference based on the feedback.

\subsection{Pseudo-Implicit Feedback}
\begin{figure}[htbp]
	\centering
	\includegraphics[width=1.85 in]{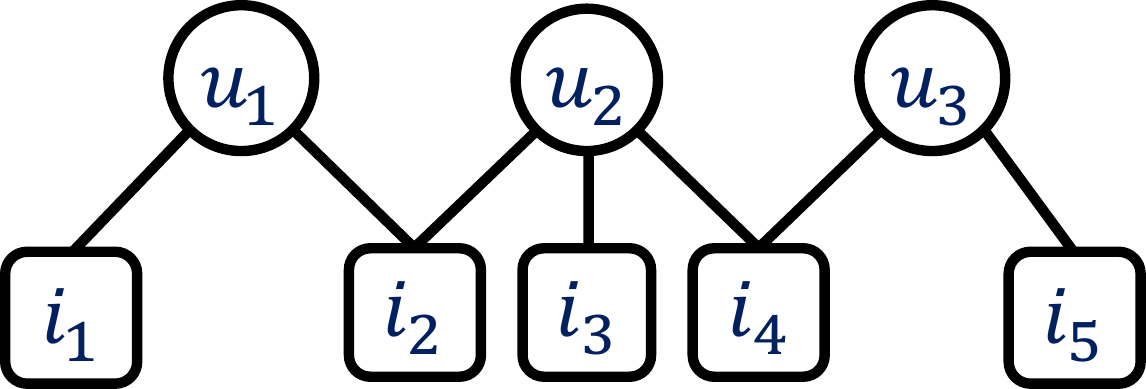}
	\caption{A toy example of the user-item bipartite graph $G$. A circle represents a user and a rectangle represents an item.}
    \label{bigraph}
\end{figure}

Complementary to viewing the user-item interactions as a matrix, we can view $\textbf{R}$ as a bipartite graph $G=(V, E)$, where $V=U \cup I$ and there exists an edge $e_{ui} \in E$ if and only if $r_{ui} = 1$. Figure \ref{bigraph} presents a toy example of $G$, where a circle represents a user and a rectangle represents an item. %We further define the adjacency matrix of $G$ as $A$.

From this perspective, we hypothesize that there are many cases where $u$ probably likes $i$ even though $u$ never purchases $i$. For example, in Figure \ref{bigraph}, $u_{1}$ and $u_{2}$ both purchased $i_{2}$, which implies that $u_{1}$ and $u_{2}$ probably share some similarities in their preference patterns. The fact that $u_{1}$ purchased $i_{1}$ indicates that $u_{1}$ probably likes $i_{1}$. Since $u_{1}$ and $u_{2}$ have similar preferences, $u_{2}$ probably likes $i_{1}$ so that he might also purchase $i_{1}$. Hence, we propose to treat these indirect relationships as evidence of \textit{pseudo-implicit feedback}. Intuitively, the more distance between the user and the item in the graph, the less confidence we have in the pseudo-implicit feedback. Formally, we define pseudo-implicit feedback as:
%Thus, it is more reasonable to let pseudo-implicit feedback be real-valued to reflect the confidence. 
%we artificially generate some user action logs, like $u_{2}$ purchasing $i_{1}$, which is referred as a \textit{pseudo-implicit feedback}. 

%Next, a formal definition of the pseudo-implicit feedback is given as:
\theoremstyle{definition}
\begin{definition}
\label{definition of s}
\textbf{\textit{Pseudo-Implicit Feedback}}. Pseudo-implicit feedback is a non-negative real-valued implicit feedback from $u$ to $i$, denoted as $s_{ui} \in \mathbb R_{\ge 0}$. The larger $s_{ui}$ is, the higher confidence on the pseudo-implicit feedback from $u$ to $i$ is.
If $u$ and $i$ are connected through a path in $G$, then $s_{ui}>0$.
The pseudo-implicit feedback matrix is denoted by $\textbf{S}$.\end{definition}

\begin{figure}[htbp]
	\centering
	\includegraphics[width=3.4 in]{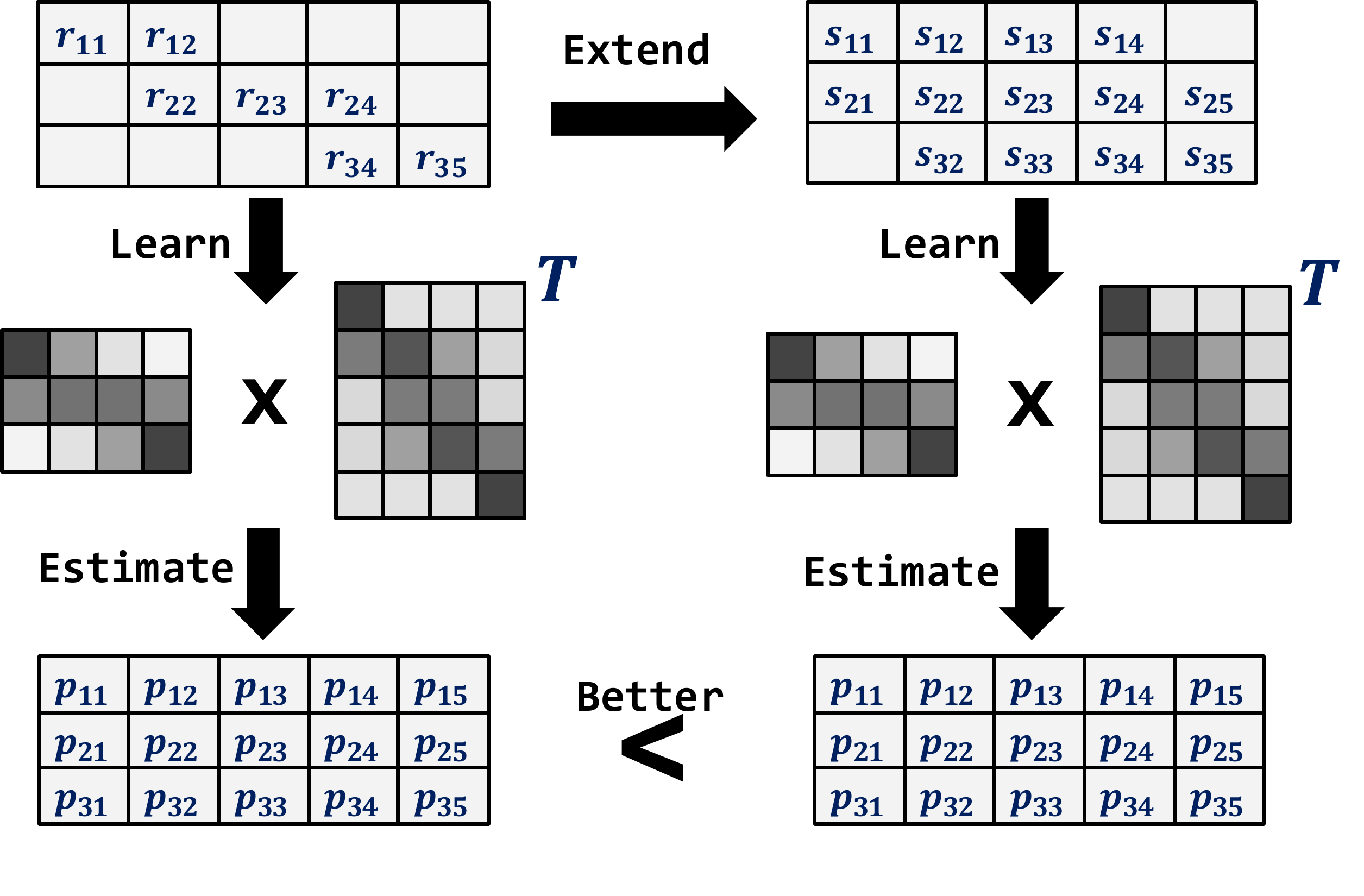}
	\caption{A toy example of generating pseudo-implicit feedback $\textbf{S}$ from original implicit feedback $\textbf{R}$ (based on Figure \ref{bigraph}) to better estimate user preference $\textbf{P}$ on very sparse datasets.}\label{overview}
\end{figure}

Figure \ref{overview} provides a toy example based on the user-item bipartite graph in Figure \ref{bigraph}. Clearly, $\textbf{S}$ is denser than $\textbf{R}$ since it also incorporates indirect user-item pairs in $G$. Then, we still follow the above-mentioned latent factor model to estimate user preferences: $\hat{p}_{ui}=x_{u}^{T}y_{i}$. The model parameters $x_{u}$ and $y_{i}$ can be learned by minimizing $\ell = \sum_{u,i}(s_{ui}-\hat{p}_{ui})^2$.
In this way, we can leverage the pseudo-implicit feedback from indirect user-item pairs to alleviate the data sparsity problem.% in top-K recommendation. 

The key question, then, is how to generate $\textbf{S}$, the pseudo-implicit feedback matrix? Care must be taken to capture meaningful (pseudo) user-item pairs without also incorporating spurious, noise-inducing user-item pairs. Hence, we propose to generate the pseudo-implicit feedback matrix by exploiting a Skip-gram inspired approach that captures user-item closeness in the bipartite graph. Specifically, we propose to estimate $\textbf{S}$ through two steps: (1) The first step is to extract user-item pairs from the graph $G$. Recall that according to  Definition \ref{definition of s}, if $i$ is reachable by $u$, then $s_{ui}>0$. Hence, we need to sample these (pseudo) user-item pairs; and (2) then measure the confidence of this feedback based on the sampled user-item pairs. Straightforward co-occurrence counts from these random walk samples is a first step, but may miss subtleties of user-item pairs not occurring in the sample. Hence, we propose a neural embedding model to overcome this challenge.

%ousince the bipartite graph $G$ the structure of the graph provides evidence of the pThus, the structure information of $G$ is the source of $S$. 

% Recall that according to Definition \ref{definition of s}, if $i$ is reachable by $u$, then $s_{ui}>0$.
 
%Then, the question can be split into two steps: (1) how to extract the structure information from $G$ and 

\begin{figure}
  \centering
  \subfigure[Generated random walks.]{
    \label{random-walks} %% label for first subfigure
    \includegraphics[width=.42\linewidth]{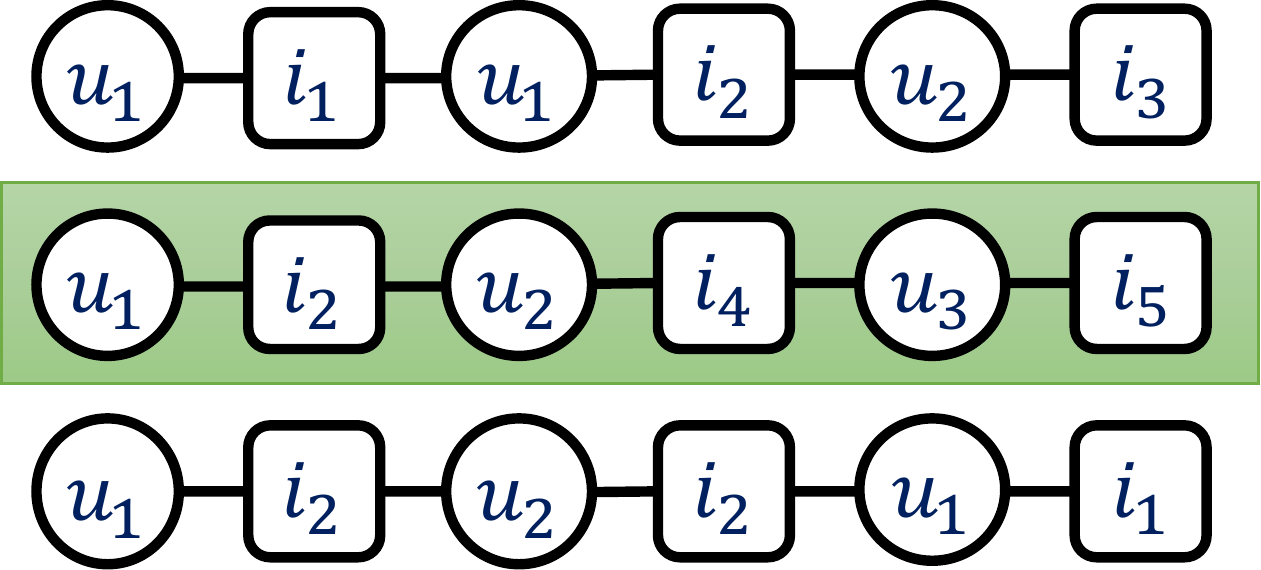}}
  \subfigure[Direct pairs.]{
      \label{direct-pairs} %% label for second subfigure
      \includegraphics[width=.3\linewidth]{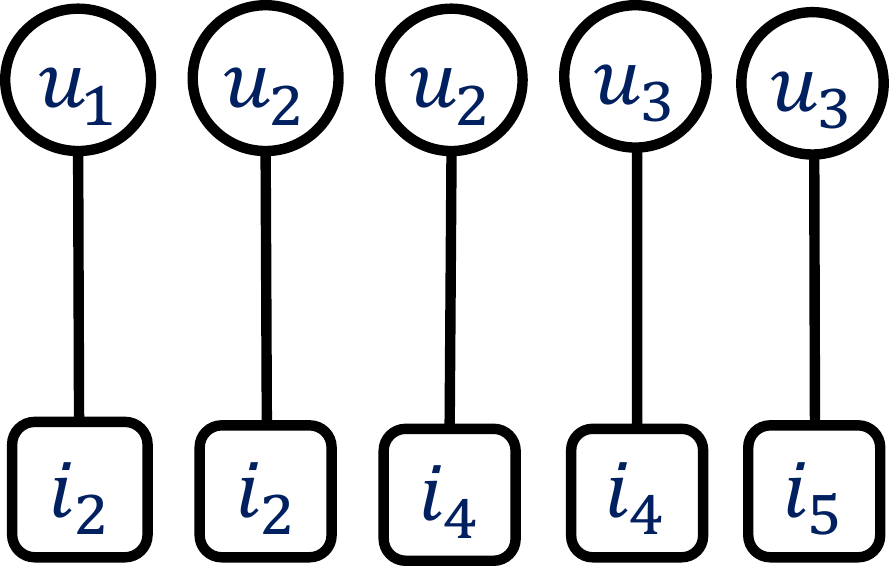}}
  \subfigure[Indirect pairs.]{
    \label{user-item-pairs} %% label for second subfigure
    \includegraphics[width=.19\linewidth]{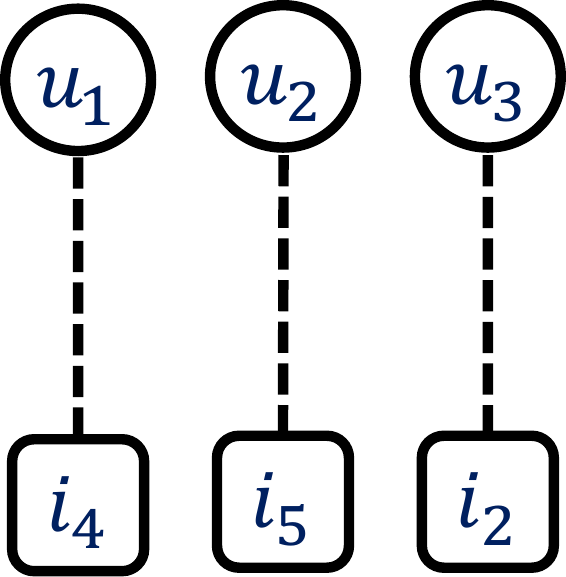}}
  \caption{Figure \ref{random-walks} shows three random walks starting from $u_{1}$ in Figure \ref{bigraph}. Figure \ref{direct-pairs} and Figure \ref{user-item-pairs} show the direct user-item pairs and indirect user-item pairs sampled from the second random walk in Figure \ref{random-walks} when the window size is 3.}
  \label{fig:sampling}
\end{figure}

%Since the information of nodes' neighborhoods in the graph $G$ is the source of pseudo-implicit feedback, o
\subsection{Random Walks to Extract User-Item Pairs}
%Our first  step is to extract connected user-item pairs from $G$. 
Inspired by previous works like DeepWalk \cite{perozzi2014deepwalk}, we propose to use truncated random walks to sample user-item pairs in the user-item bipartite graph. Recall that a random walk is a stochastic process with random variables.
Concretely, a truncated random walk can be represented as a linear sequence of vertices $v_{i}$, $v_{i+1}$,...,$v_{i+\gamma}$, where ${v_{i+k}}$ is a vertex sampled uniformly from the neighbors of ${v_{i+k -1}}$.
For each vertex $v_{i} \in V=U \cup I$, we launch $\beta$ walks of length $\gamma$ on $G$, forming the set of random walks $\mathcal{W}$.
We use $\mathcal{W}_{v_{i}}$ to represent a random walk starting from vertex $v_{i}$.

%\begin{algorithm}%[!h]
%	\label{alg:sample-rw}
%  \caption{Generate random walks from the bipartite graph $G(V, E)$}%
%	\hspace*{\algorithmicindent} \textbf{Input:} $G(V,E)$, walks per vertex $\beta$, walk length $\gamma$\\
%    \hspace*{\algorithmicindent} \textbf{Output:} Random walks $\mathcal{W}$.
%    \begin{algorithmic}[1]%
%    \STATE $\mathcal{W}=\emptyset$
%    \FOR {\textbf{each} i=0 to $\beta$}
%    \FOR {\textbf{each} $v_{i} \in V$}
%    \STATE $\mathcal{W}_{v_{i}} = RandomWalk(G,v_{i},\gamma)$
%    \STATE $\mathcal{W} = \mathcal{W} \cup \mathcal{W}_{v_{i}}$
%    \ENDFOR
%    \ENDFOR
%    \end{algorithmic}
%\end{algorithm}
%First, Algorithm 1 presents how we generate random walks from the user-item bipartite graph. 

%In our preliminary experiments, we considered random walks with restart; that is, where each walk could return to the starting vertex with a teleportation probability. In this way, random walks with different lengths can be obtained depending on when the random walker teleports back to the starting vertex. However, the restart random walks achieve no significant improvement in our experiments. Therefore, we fix the random walk length $\gamma$ in our presentation and in the experiments reported in this paper.

\begin{algorithm}[!h]
	\label{Sample Pair}
    \caption{Sample $(u,i)$ from the random walks $\mathcal{W}$}%
	\hspace*{\algorithmicindent} \textbf{Input:} User set $U$, item set $I$, random walks $\mathcal{W}$, window size $\sigma$\\
    \hspace*{\algorithmicindent} \textbf{Output:} The set of $(u,i)$.
    \begin{algorithmic}[1]
    \STATE $C=\emptyset$
    \FOR {\textbf{each} $\mathcal{W}_{v_{i}} \in \mathcal{W}$ }
    \FOR {\textbf{each} $\mathcal{W}_{v_i}^j \in \mathcal{W}_{v_i}$ and $\mathcal{W}_{v_i}^j \in U$}
    \FOR {\textbf{each} $k \in [j-\sigma:j+\sigma:2]$}
    % \FOR {each $v_{k} \in A_{v_{i}}[j-\sigma:j+\sigma]$ and $v_{k} \in I$}
    \STATE $C = C \cup (\mathcal{W}_{v_i}^j, \mathcal{W}_{v_i}^k)$
    \ENDFOR
    \ENDFOR
    \ENDFOR
    \end{algorithmic}
\end{algorithm}

%Our algorithm for sampling user-item pairs is presented in Algorithm 1. 
Once the corpus of random walks $\mathcal{W}$ is obtained, we sample user-item pairs from it as presented in Algorithm 1. In Lines 2-3, we iterate over all users in all random walk sequences. For each user vertex ($v \in U$) in the sequences, any item vertex ($v \in I$) whose distance from $v$ is not greater than the window size will be sampled. An example of the output of Algorithm 1 is shown in Figure \ref{random-walks}, where we sample three random walks starting from $u_1$ in the graph presented in Figure \ref{bigraph}.
Figure \ref{direct-pairs} and Figure \ref{user-item-pairs} show the direct and indirect user-item pairs sampled from the second random walk in Figure \ref{random-walks} when the window size is set to 3.

\subsection{Measuring Pseudo-Implicit Feedback}
The next question is how to measure the pseudo-implicit feedback $s_{ui}$ from the sampled user-item pairs above.
% Let $F$ be the function of $s_{ui}$ such that we have: $s_{ui}=F(u,i)$ where the number of possibilities of $F$ is vast.
% In this subsection, we investigate $F$ by two strategies:
In this subsection, we investigate two strategies for estimating $s_{ui}$ for each $(u, i)$ pair:
co-occurrence counts and pointwise mutual information (PMI) \cite{church1990word}. Our approach using the two strategies are denoted as \textit{PsiRec-CO} and \textit{PsiRec-PMI} respectively. We first discuss the co-occurrence counts strategy. After that, to overcome its limitation, the latter one is presented as well as its connection to the well-known Skip-gram embedding model. The comparison of their performance will be discussed in Section \ref{comparison of variants}.

\subsubsection{Co-occurrence counts}
Let $(u,i)$ represents a pair of $u$ and $i$. One straightforward measurement of $s_{ui}$ is the total number of times $(u,i)$ appear in $C$:
\begin{equation}
	 s_{ui}=\# (u, i)
\end{equation}

However, this measurement suffers from the drawback that $\# (u, i)$ is strongly influenced by the popularity (i.e. vertex degree) of users and items. As a consequence, for most users high degree items will dominate their recommendations \cite{christoffel2015blockbusters}.
Similarly, for most items, their co-occurrence count with the high degree users will be large.
The deeper reason behind this drawback is that it only considers the positive samples -- the user-item pairs appearing in $C$, but ignores the negative samples -- the user-item pairs not appearing in $C$. %Intuitively, $\# (u, i)$ should be normalized by $\# (u)$ and $\# (i)$, that are the number of times $u$ and $i$ appear in $C$ respectively, such that:
%\begin{equation}
%\label{Intuition of F}
%	s_{ui}=F(\frac{\# (u, i)}{\# (u) \cdot \# (i)})
%\end{equation}
%where $F$ is a non-decreasing function.

\subsubsection{Pointwise mutual information}
To overcome this limitation, we first build an embedding model to measure $s_{ui}$ and then further propose an improved version of the model which explicitly factorizes a matrix where each entry is the pointwise mutual information (PMI) value of $u$ and $i$ in $C$. Motivated by the Skip-gram with negative sampling (SGNS) model \cite{mikolov2013distributed},
we aim to maximize the probability of observing the user-item pairs that appear in the corpus $C$
and minimize the probability of observing pairs that do not appear in $C$. We denote the embedding vector for user $u$ as $\vec{u}$ and
the embedding vector for item $i$ as $\rarrow{i}$.
% Let $P(D = 1 | u, i)$ be the probability that $(u, i)$ is from $C$, and
% $P(D = 0 | u, i) = 1 - P(D = 1 | u, i)$ the probability that $(u, i)$ is not.
The probability that $(u, i)$ is from $C$ is:
$$P(D = 1 | u, i) = \sigma (\rarrow{u} \cdot \rarrow{i}) =
\frac{1}{1 + e^{-\rarrow{u} \cdot \rarrow{i}}}$$

For each positive example, we then draw several random items $i_N$ from the set of items $I$
as negative examples.
Since the randomly drawn items are unlikely to co-occur with $u$,
we want to minimize the probability of observing these $(u, i_N)$ pairs.
The training objective for a single $(u, i)$ observation is:
\begin{equation}
  \log \sigma(\rarrow{u} \cdot \rarrow{i}) + k \cdot \mathbb{E}_{i_N \sim P_D}
  [\log \sigma(-\rarrow{u} \cdot \rarrow{i_N})]
\end{equation}
where $k$ is the number of negative samples and $i_N$ is a random item drawn according to
a distribution $P_D$ over $I$.
Here, following SGNS, we use $P_D = \frac{\# (i)}{|C|}$
where $\# (i)$ is the number of occurrences of $i$ in $C$.
The global training objective can be defined over all $(u, i) \in C$:
\begin{equation}
  \label{eq:sgns-obj}
  \ell = \sum_{(u, i) \in C}(\log \sigma(\rarrow{u} \cdot \rarrow{i}) + k \cdot \mathbb{E}_{i_N \sim P_D}
  [\log \sigma(-\rarrow{u} \cdot \rarrow{i_N})])
\end{equation}
which can be optimized with stochastic gradient descent. %and the pseudo-implicit feedback can be measured by: $s_{ui}=\rarrow{u} \cdot \rarrow{i}$.

However, the above embedding model fails to take advantage of the large amount of repetitions in $C$.
A user-item pair may have hundreds or thousands of occurrences in the corpus,
incurring a large time cost for model training.
Thus, a natural question to ask is: can we directly operate on the co-occurrence statistics
of the user-item corpus? To answer this question, inspired by \cite{levy2014neural}, we first re-write the loss function in Eq. \ref{eq:sgns-obj} w.r.t each unique
user-item pair:
\begin{equation}
  \label{eq:sgns-1}
  \begin{array}{l}
  \ell = \sum_{u \in U}\sum_{i \in I} \# (u, i) (\log \sigma(\rarrow{u} \cdot \rarrow{i}) + \\
  k \cdot \# (u) \cdot \mathbb{E}_{i_N \sim P_D}
  [\log \sigma(-\rarrow{u} \cdot \rarrow{i_N})])
  \end{array}
\end{equation}
Additionally, we explicitly express the expectation term $\mathbb{E}_{i_N \sim P_D}
[\log \sigma(-\rarrow{u} \cdot \rarrow{i_N})]$ as follows:
\begin{equation}
  \label{eq:sgns-2}
  \begin{array}{l}
  \mathbb{E}_{i_N \sim P_D}[\log \sigma(-\rarrow{u} \cdot \rarrow{i_N})] = \\
  \frac{\# (i)}{|C|}   \log \sigma(-\rarrow{u} \cdot \rarrow{i}) + 
  \sum\limits_{i_N \in I \backslash \{i\}} \frac{\# (i_N)}{|C|} \log \sigma(-\rarrow{u} \cdot \rarrow{i_N})
  \end{array}
\end{equation}
Combining Eq. \ref{eq:sgns-1} and Eq. \ref{eq:sgns-2} leads to a local loss for a specific user-item pair $(u, i)$:
\begin{equation}
\label{eq:sgns-3}
\ell(u, i) = \# (u, i) \log \sigma(\rarrow{u} \cdot \rarrow{i}) +
k \cdot \# (u) \cdot \frac{\# (i)}{|C|} \log \sigma(-\rarrow{u} \cdot \rarrow{i})
\end{equation}

Let $s^\prime_{ui} = \rarrow{u} \cdot \rarrow{i}$. We then take the derivative of Eq. \ref{eq:sgns-3} w.r.t. $s^\prime_{ui}$ and set it
to zero, which eventually leads to:
\begin{equation}
\label{PMI value}
s^\prime_{ui} = \rarrow{u} \cdot \rarrow{i} = \log (\frac{\# (u, i) \cdot |C|}{\# (u) \cdot \# (i)}) - \log k
\end{equation}
We note that the term $\log (\frac{\# (u, i) \cdot |C|}{\# (u) \cdot \# (i)})$ is the pointwise mututal information (PMI) of $(u, i)$, which is widely used to measure the association between random variables.
This indicates that the objective function in Eq. \ref{eq:sgns-obj} is implicitly factorizing a user-item association matrix, where each entry is the PMI of $(u, i)$ shifted by a constant $\log k$.

However, we notice that the above formulation of $s^\prime_{ui}$ has a problem: for $(u, i)$ pairs that never appear in $C$, their PMI are ill-defined since $\# (u, i) = 0$. Moreover, it is suggested that $(u, i)$ pairs with very low or even negative PMI scores are less informative and should be discarded \cite{bullinaria2007extracting}.
Thus, we propose to compute $s_{ui}$ by applying the following transformation:
\begin{equation}
\label{s equation}
  s_{ui} = \max (s^\prime_{ui}, 0)
\end{equation}
This formulation of $s_{ui}$ assigns zeros to both $(u, i)$ pairs that do not appear in $C$ and those with very low or negative PMI values,
thus, solving both problems above.
In addition, $\textbf{S}$ is a non-negative matrix which is consistent with Definition \ref{definition of s}.

\subsection{Latent Factor Model to Estimate User Preference}
Finally, we propose a matrix factorization approach to estimate $p_{ui}$ based on the learned pseudo-implicit feedback matrix $S$, where the loss function is:
%Latent factors are learned by minimizing the loss function in Eq. \ref{eq:wmf-loss}:
\begin{equation}
\label{eq:wmf-loss}
	\ell = \sum_{u,i}(s_{ui}-\textbf{x}_{u}^{T}\textbf{y}_{i})^2+\lambda(\sum_{u}|\vert \textbf{x}_{u} |\vert ^{2} + \sum_{i}|\vert \textbf{y}_{i} |\vert ^{2})
\end{equation}
where $\lambda$ controls the strength of regularization to prevent overfitting. Inspired by \cite{hu2008collaborative}, we apply the well-known Alternating Least Square (ALS) algorithm to optimize the loss function in Eq. \ref{eq:wmf-loss}.
Let $\textbf{s}_{u} \in \mathbb{R}^{N}$ represents the pseudo-implicit feedback over all items of $u$ and $\textbf{s}_{i} \in \mathbb{R}^{M}$ represents the pseudo-implicit feedback of $i$ over all users.
By setting $\frac{\partial \ell}{\partial \textbf{x}_{u}}=0$ and $\frac{\partial \ell}{\partial \textbf{y}_{i}}=0$, we have:

\begin{equation}
\label{x}
\textbf{x}_{u} = (\textbf{Y}^{T}\textbf{Y}+\lambda \textbf{D})^{-1}\textbf{Y}^{T}\textbf{s}_{u}
\end{equation}

\begin{equation}
\label{y}
\textbf{y}_{i} = (\textbf{X}^{T}\textbf{X}+\lambda \textbf{D})^{-1}\textbf{X}^{T}\textbf{s}_{i}
\end{equation}

\noindent where $\textbf{D}$ represents the identity matrix.
After $\textbf{X}$ and $\textbf{Y}$ are learned, the whole predicted preference matrix $\hat{\textbf{P}} \in \mathbb{R}^{M \times N}$ is reconstructed by $\hat{\textbf{P}}=\textbf{X}^{T}\textbf{Y}$, where $\hat{p}_{ui}$ is used to rank items.

\section{Experimental Evaluation}
% Our experiments focus on three research questions:% by analyzing the experimental results:

% \begin{enumerate}
% 	\item Which strategy is the best to measure $s_{ui}$?
% 	\item Does our proposed approach outperform the strong baselines, especially the two start-of-the-art neural network methods?
% 	\item How well does our approach alleviate the data sparsity problem for the implicit recommendation task?
	%\item How does the choices of hyperparameters affect the performance of PsiRec?
%\end{enumerate}
%\label{Experimental Evaluation}
%In this section, we investigate the impact of our pseudo-implicit feedback framework on recommendation quality under extreme sparsity. We first introduce three publicly accessible datasets. Then we present evaluation metrics and five baseline methods. We investigate recommendation quality versus these baselines, explore the impact of sparsity (by varying the number of interactions available to the method), and evaluate how parameters (like the window size and number of random walks) affect the performance of our approach.

\subsection{Datasets and Evaluation Metrics}
\label{data sets}
%We evaluate our approach on two public datasets demonstrating user-item interaction sparsity: Amazon Toys and Games,  and the Tmall shopping logs dataset from the IJCAI Challenge 2015\footnote{https://tianchi.aliyun.com/datalab/dataSet.html? spm=5176.100073.0.0.288635eeRA3jPq\&dataId=47}. %Amazon Cell Phones and Accessories

\medskip
\noindent\textbf{Amazon.} The Amazon dataset\footnote{http://jmcauley.ucsd.edu/data/amazon/links.html} contains explicit feedback, where users give a rating from 1 to 5 on the purchased items. Following previous work \cite{hu2008collaborative,ncf}, we transform these explicit datasets into implicit datasets by treating all ratings as 1, indicating that the user has purchased the item. 

\medskip
\noindent\textbf{Tmall.} The Tmall dataset\footnote{https://tinyurl.com/y722nrgu} contains anonymized shopping logs of customers on Tmall.com. Originally, the dataset consists of different types of user implicit feedback, such as clicks, add-to-cart actions, and purchases. For our experiments, we focus on the purchase action logs and transform them into the binary interactions $\textbf{R}$ since these are closely aligned with our purchase prediction task. Further, we filter out users and items with fewer than 20 interactions.

\medskip
Table \ref{data statistics} summarizes the basic statistics of these two datasets. Note that the densities of the two datasets are 0.072\% and 0.051\% respectively. We also observe that even after performing the filtering on the Tmall dataset described above that it is still the sparser one.

For our experiments, we randomly split each dataset into three parts: 80\% for training, 10\% for validation (for parameter tuning only) and 10\% for testing (for evaluation). Since many users have very few interactions, they may not appear in the testing set at all. In this case, there will be no match between the top-K recommendations of any algorithm and the ground truth since there are no further purchases made by these users. The results reported here include these users, reflecting the challenges of recommendation under extreme sparsity.%. For them, no interactions are randomly selected into the testing part. Specifically, the number of users are 10429, 13812 and 20876 on the testing part of Amazon Toys, Amazon Cell Phones and Tmall dataset respectively. Although top-K recommendation results of them will be zero no matter which metrics are used, they are still incorporated into the calculation of the metrics to reflect the suffering of all methods on very sparse datasets.  %Moreover, each user or item appears at least once in the three parts.  
%\textbf{For instance, for each interaction, we draw a number from ${\rm Uniform}(0,1)$ and if the number is smaller than 0.1, then it will be selected into the testing part. 

% Table generated by Excel2LaTeX from sheet 'Data Stat'
\begin{table}[htbp]
  \centering
	\setlength{\tabcolsep}{1.3pt}
  \caption{Summary statistics for the evaluation datasets}
    \begin{tabular}{lccccc}
	\toprule
    Dataset & Users & Items & Interactions& Density  & $\frac{Interactions}{User}$\\
	\midrule
    Amazon Toys & 19,412 & 11,924 & 167,597 & 0.072\% & 8.63\\
    %Amazon Cell Phones & 27,879 & 10,429 & 194,439 & 0.067\% & 6.97\\
    Tmall & 28,291 & 28,401 & 410,714 & 0.051\% & 14.51\\
	\bottomrule
    \end{tabular}%
  \label{data statistics}%
\end{table}%
 
%\subsection{Evaluation Metrics}
\noindent\textbf{Metrics.} Given a user, a top-K item recommendation algorithm provides a ranked list of items according to the predicted preference for each user. To assess the ranked list with respect to the ground-truth item set of what users actually purchased, we adopt three evaluation metrics: precision@k, recall@k, and F1@k, where we average each metric across all users.

%including precision at 5 (P@5), 10 (P@10), recall at 5 (R@5), 10 (R@10) and F1 value at 5 (F1@5), 10 (F1@10). 

%Precision@K measures the fraction of correctly predicted items among the top-K recommended items:
%
%\begin{equation}
%P@K=\sum_{i=1}^{K}\frac{R(K)}{K}
%\end{equation}
%where $R(i)$ is an indication function of the \textit{i}-th item, $R(i)=1$ indicates the \textit{i}-th item is purchased by the user and 0 otherwise.
%
%Recall measures the fraction of correctly predicted items among all the items purchased by the user:
%
%\begin{equation}
%R@K=\sum_{i=1}^{K}\frac{R(K)}{N(u)}
%\end{equation}
%where $N(u)$ is the total number of items purchased by user $u$ in the testing dataset.
%
%The F1 value is the harmonic mean of precision and recall:
%
%\begin{equation}
%F1@K=\frac{2 \cdot P@K \cdot R@K}{P@K+R@K}
%\end{equation}

\subsection{Baselines}
%We compare our approach with several top-K recommendation methods, including standard matrix factorization techniques and two state-of-the-art neural network methods:

\medskip
\noindent\textbf{ItemPop}. This simple method recommends the most popular items to all users, where items are ranked by the number of times they are purchased. %The same top-K most popular items are recommended to all users. 

\medskip
\noindent\textbf{BPR} \cite{rendle2009bpr}. Bayesian personalized ranking (BPR)  is a standard pairwise ranking framework for implicit recommendation. %BPR assumes that, for each user, the preference of the observed item ($r_{ui}=1$) is superior to the non-observed ones ($r_{ui}=0$). 

\medskip
\noindent\textbf{MF} \cite{koren2009matrix}. Traditional matrix factorization (MF) method uses mean squared error as the objective function with negative sampling from the non-observed items ($r_{ui}=0$). 

\medskip
\noindent\textbf{NCF} \cite{ncf}. Neural collaborative filtering (NCF) is a state-of-the-art neural network recommender. NCF concatenates latent factors learned from a generalized matrix factorization model and a multi-layered perceptron model and then use a regression layer to predict user preferences. %Following \cite{ncf}, negative samples are uniformly sampled from unobserved interactions ($r_{ui}=0$). The number of negative samples is tuned as a hyperparameter over the validation set.% on the testing dataset.

\medskip
\noindent\textbf{CDAE} \cite{cdae}. Collaborative Denoising Auto-Encoders (CDAE) is another successful neural network approach for top-K recommendation. A standard Denoising Auto-Encoder takes $\textbf{R}$ as input and reconstructs it with an auto-encoder. CDAE extends the standard Denoising Auto-Encoder by also encoding a user latent factor. %Finally, items are ranked by the reconstructed interactions matrix.

%\subsection{Variants of Our Model}

%\subsubsection{Random Walks}
%\subsection{Analysis of Random Walks}
%In this work, 

%\subsubsection{Neural Embedding Learning}
%Following \cite{perozzi2014deepwalk}, random walks are used to generate linear sequences of vertices from the user-item bipartite graph. Then skipgram is used to learn the dense embedding vectors of the vertices. We calculate the cosine similarity between the embeddings of the user and the item and the items are ranked by the similarities.

\subsection{Hyperparameter Settings}
Our preliminary experiments show that a large number of latent factors can be helpful for learning user preferences on sparse datasets. Hence, for all methods, we set the number of latent factors to 100. For all baseline methods, we carefully choose the other hyperparameters over the validation set.%by the validation on the testing dataset. %We randomly initialized model parameters with a Gaussian distribution (with a mean of 0 and standard deviation of 0.01). 

Particularly, for our approach, the regularization parameter $\lambda$ is 0.25, the number of random walks $\beta$ for each node is 10. The length $\gamma$ of each random walk is 80. The window size $\sigma$ for sampling the user-item pairs is 3. %The impact of these parameters is discussed below. %Shifted value $k$ is 3.

%\section{Experimental Results}
%\label{Experimental Results}

%\subsection{Effectiveness of Random Walks}
%\subsection{Effectiveness of Our Latent Factors Model}
\subsection{Comparison between the Two Strategies for Measuring $\textbf{S}$}
\label{comparison of variants}
We first compare the performance of \textit{PsiRec-CO} and \textit{PsiRec-PMI}. The results in Table \ref{Toy results}, \ref{Tmall results} show that PMI is significantly superior to co-occurrence as the measure of pseudo-implicit feedback. The higher the degree of a vertex in the user-item bipartite graph, the higher probabilities of the vertex co-occurring with other vertices. Hence, a high $\# (u, i)$ might not guarantee a high confidence if $\# (i)$ or $\# (u)$ is also very large. Therefore, it is more reasonable to let $\# (u, i)$ to be normalized by the global frequency of $u$ and $i$, which is exactly what PMI captures. Since \textit{PsiRec-PMI} brings considerable improvements upon \textit{PsiRec-CO}, we focus on \textit{PsiRec-PMI} for the rest of this paper.

%% Table generated by Excel2LaTeX from sheet 'PPMI'
%\begin{table}[htbp]
%  \centering
%\setlength{\tabcolsep}{2.6pt}
%  \caption{Comparison between $PPMI(u,i)$ and $o(u,i)$}
%    \begin{tabular}{|c|cccccc|}
%	\toprule
%       Metric(\%)   & \multicolumn{1}{l}{P@5} & \multicolumn{1}{l}{R@5} & \multicolumn{1}{l}{F1@5} & \multicolumn{1}{l}{P@10} & \multicolumn{1}{l}{R@10} & F1@10\\
%	\midrule
%	WMF	& 0.533 & 1.716 & 0.813 & 0.403 & 2.532 & 0.695 \\
%    Model II $o(u,i)$    & 0.633 & 2.100 & 0.972 & 0.495 & 3.249 & 0.859 \\
%    Model II $PMI(u,i)$  & $0.955^{*}$ & $3.224^{*}$ & $1.474^{*}$ & $0.728^{*}$ & $4.824^{*}$ & $1.265^{*}$ \\
%	\midrule
%    Improvement & +51.0 & +53.5 & +51.6 & +47.0 & +48.5 & +47.2 \\
%	\bottomrule
%    \end{tabular}%
%  \label{PPMI and RC}%
%\end{table}%

\subsection{Comparison with Baselines}
Tables \ref{Toy results}, \ref{Tmall results} also show the experimental results of the baseline methods, along with the improvement of \textit{PsiRec-PMI} over the strongest baseline CDAE. Statistical significance was tested by the t-test, where ``*'' denotes statistical significance at $p = 0.01$ level.

On both datasets, we observe that \textit{PsiRec-PMI} significantly outperforms all baseline methods on all metrics. For instance, \textit{PsiRec-PMI} achieves the best P@10 of 0.728\% and 0.978\% on the two datasets, representing 20.5\% and 26.8\% improvement upon the strongest baseline method CDAE.
Similarly, significant improvements in terms of recall are also observed.
\textit{PsiRec-PMI} achieves this by combining pseudo-implicit feedback with a simple matrix factorization model, which is much more scalable and easier to tune than the state-of-the-art neural network methods.
The experimental results demonstrate that, on sparse datasets, matrix factorization method largely benefits from considering indirectly transitive user-item relationships, and pseudo-implicit feedback generated by \textit{PsiRec-PMI} can better estimate the users' actual preferences. %Besides, the worst performance obtained by \textit{ItemPop} indicates that users' preference is considerably diverse. This might be a characteristic of online store datasets (e.g. Amazon, Tmall), which is different with datasets where popular culture dominates users' preference greatly, as in music and movie domains.%, like music dataset \footnote{https://labrosa.ee.columbia.edu/millionsong/}. Surprisingly, BPR achieves the second worst results in our experiments. The preference information from indirect user-item pairs is not utilized in this method. Therefore, some of BPR's failures might stem from ignoring the difference among users' preference upon non-observed items.

% Table generated by Excel2LaTeX from sheet 'All'
\begin{table}[htbp]
  \centering
	\setlength{\tabcolsep}{2.2pt}
	\renewcommand\arraystretch{0.8}
  \caption{Experimental Results on Amazon Toys and Games}
    \begin{tabular}{ccccccc}
	\toprule	
    Metric(\%) & \multicolumn{1}{l}{P@5} & \multicolumn{1}{l}{R@5} & \multicolumn{1}{l}{F1@5} & \multicolumn{1}{l}{P@10} & \multicolumn{1}{l}{R@10} & F1@10 \\
	\midrule
    ItemPop & 0.127 & 0.412 & 0.194 & 0.112 & 0.753 & 0.195 \\
    BPR   & 0.183 & 0.616 & 0.283 & 0.159 & 1.087 & 0.277 \\
    MF    & 0.533 & 1.716 & 0.813 & 0.403 & 2.532 & 0.695 \\
    NCF   & 0.733 & 2.529 & 1.136 & 0.531 & 3.565 & 0.924 \\
    CDAE  & 0.801 & 2.727 & 1.238 & 0.604 & 4.044 & 1.051 \\
    \midrule
    PsiRec-CO  & 0.633 & 2.100 & 0.972 & 0.495 & 3.249 & 0.859 \\
    %Model I & 0.433 & 1.467 & 0.668 &	0.333 & 2.175 & 0.578 \\
    PsiRec-PMI   & $0.955^{*}$ & $3.224^{*}$ & $1.474^{*}$ & $0.728^{*}$ & $4.824^{*}$ & $1.265^{*}$ \\
	\midrule
	Improvement(\%)	& +19.3 & +18.2 & +19.1 & +20.5	& +19.3	 & +20.3\\
	\bottomrule
    \end{tabular}%
  \label{Toy results}%
\end{table}%
	
% Table generated by Excel2LaTeX from sheet 'All'
%\begin{table}[htbp]
%  \centering
%	\setlength{\tabcolsep}{2.4pt}
%	\renewcommand\arraystretch{0.9}
%  \caption{Experimental Results on Amazon Cell Phones Dataset}
%    \begin{tabular}{|c|cccccc|}
%	\toprule
%    Metric(\%) & \multicolumn{1}{l}{P@5} & \multicolumn{1}{l}{R@5} & \multicolumn{1}{l}{F1@5} & \multicolumn{1}{l}{P@10} & \multicolumn{1}{l}{R@10} & F1@10 \\
%	\midrule
%    ItemPop & 0.251 & 0.920 & 0.395 & 0.219 & 1.636 & 0.386 \\
%    BPR   & 0.289 & 1.045 & 0.453 & 0.238 & 1.734 & 0.419 \\
%    MF    & 0.533 & 1.915 & 0.834 & 0.414 & 2.927 & 0.725 \\
%    NCF   & 0.613 & 2.250 & 0.964 & 0.442 & 3.205 & 0.777 \\
%    CDAE  & 0.807 & 2.982 & 1.270 & 0.608 & 4.448 & 1.069 \\
%    %Model I & 0.297 & 1.077 &	0.466 & 0.234 & 1.716 & 0.412\\
%    \midrule
%    PsiRec-CO & 0.709 & 2.60 & 1.11 & 0.534 & 3.88 & 0.938\\
%    PsiRec-PMI   & $0.983^{*}$ & $3.633^{*}$ & $1.547^{*}$ & $0.738^{*}$ & $5.457^{*}$ & $1.300^{*}$ \\
%	\midrule
%	Improvement(\%) & +21.8 & +21.8 & +21.8 & +21.5 & +22.7 & +21.6\\
%	\bottomrule
%    \end{tabular}%
%  \label{Cell results}%
%\end{table}%
	
% Table generated by Excel2LaTeX from sheet 'All'
\begin{table}[htbp]
  \centering
	\setlength{\tabcolsep}{2.2pt}
	\renewcommand\arraystretch{0.8}
  \caption{Experimental Results on Tmall}
    \begin{tabular}{ccccccc}
	\toprule	
    Metric(\%) & \multicolumn{1}{l}{P@5} & \multicolumn{1}{l}{R@5} & \multicolumn{1}{l}{F1@5} & \multicolumn{1}{l}{P@10} & \multicolumn{1}{l}{R@10} & F1@10 \\
	\midrule
    ItemPop & 0.168 & 0.436 & 0.243 & 0.150 & 0.775 & 0.251 \\
    BPR   & 0.245 & 0.574 & 0.344 & 0.209 & 0.965 & 0.343 \\
    MF    & 1.010 & 2.466 & 1.433 & 0.646 & 3.124 & 1.071 \\
    NCF   & 0.747 & 1.914 & 1.074 & 0.537 & 2.730 & 0.897 \\
    CDAE & 1.126 & 2.877 &	1.619 & 0.771 & 3.955 & 1.291 \\
    %Model I & 0.638 & 1.694 &	0.927 & 0.436 & 2.279 & 0.732
    \midrule
    PsiRec-CO & 1.142 & 2.773 & 1.618 & 0.732 & 3.556 & 1.214 \\
    PsiRec-PMI   & $1.444^{*}$ & $3.663^{*}$ & $2.072^{*}$ & $0.978^{*}$ & $4.952^{*}$ & $1.633^{*}$ \\
	\midrule
	Improvement(\%)	&+28.2&	+27.3&	+28.0&	+26.8&	+25.2&	+26.5\\
	\bottomrule
    \end{tabular}%
  \label{Tmall results}%
\end{table}%

\begin{figure*}
  \centering
  \subfigure[Interactions Kept: $100\%$]{
    \label{100} %% label for first subfigure
    \includegraphics[width=1.3in]{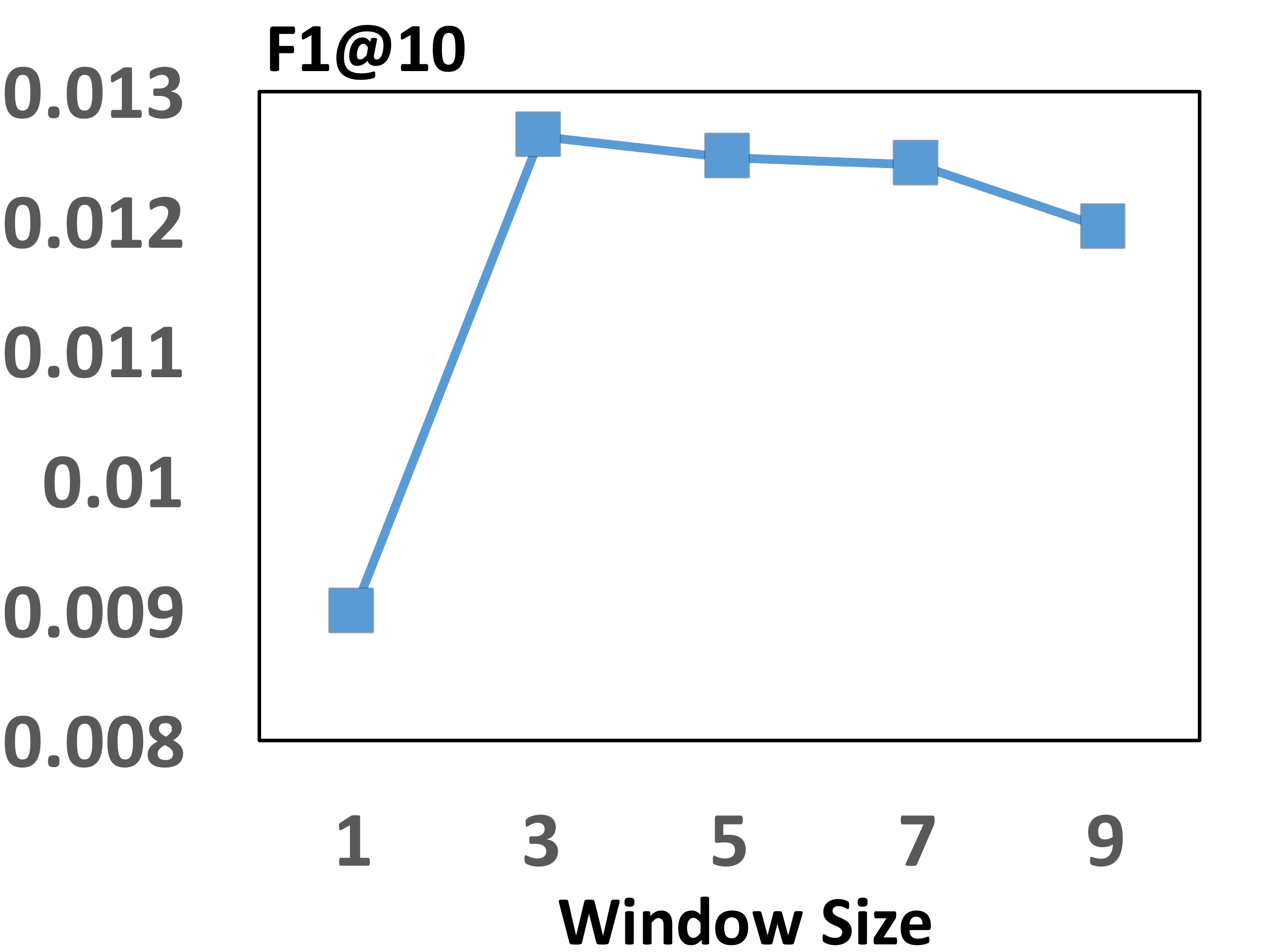}}
  %\hspace{1in}
  \subfigure[Interactions Kept: $80\%$]{
    \label{80} %% label for second subfigure
    \includegraphics[width=1.3in]{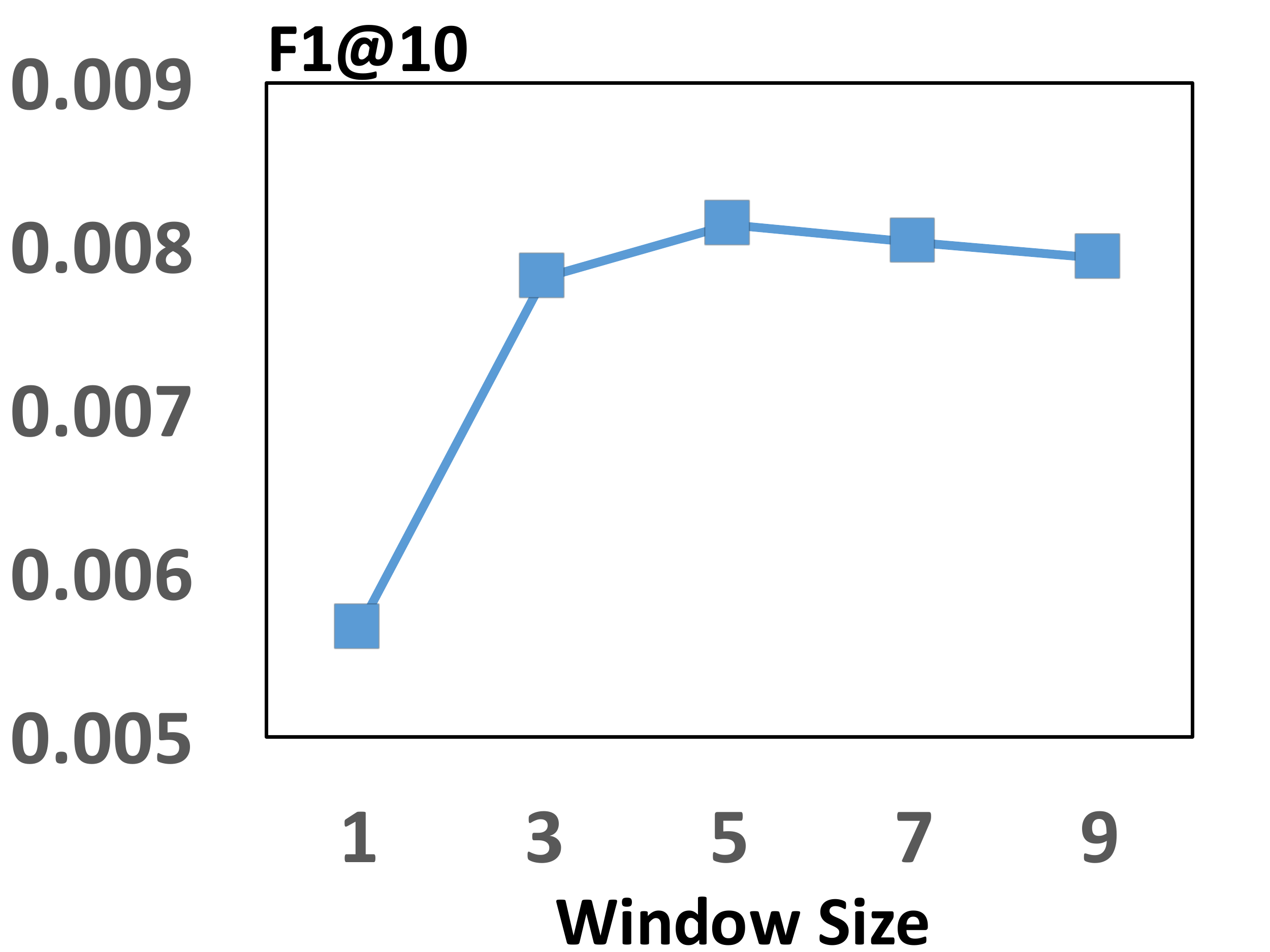}}
  %\hspace{1in}
 %% label for entire figure
  \subfigure[Interactions Kept: $60\%$]{
    \label{60} %% label for second subfigure
    \includegraphics[width=1.3in]{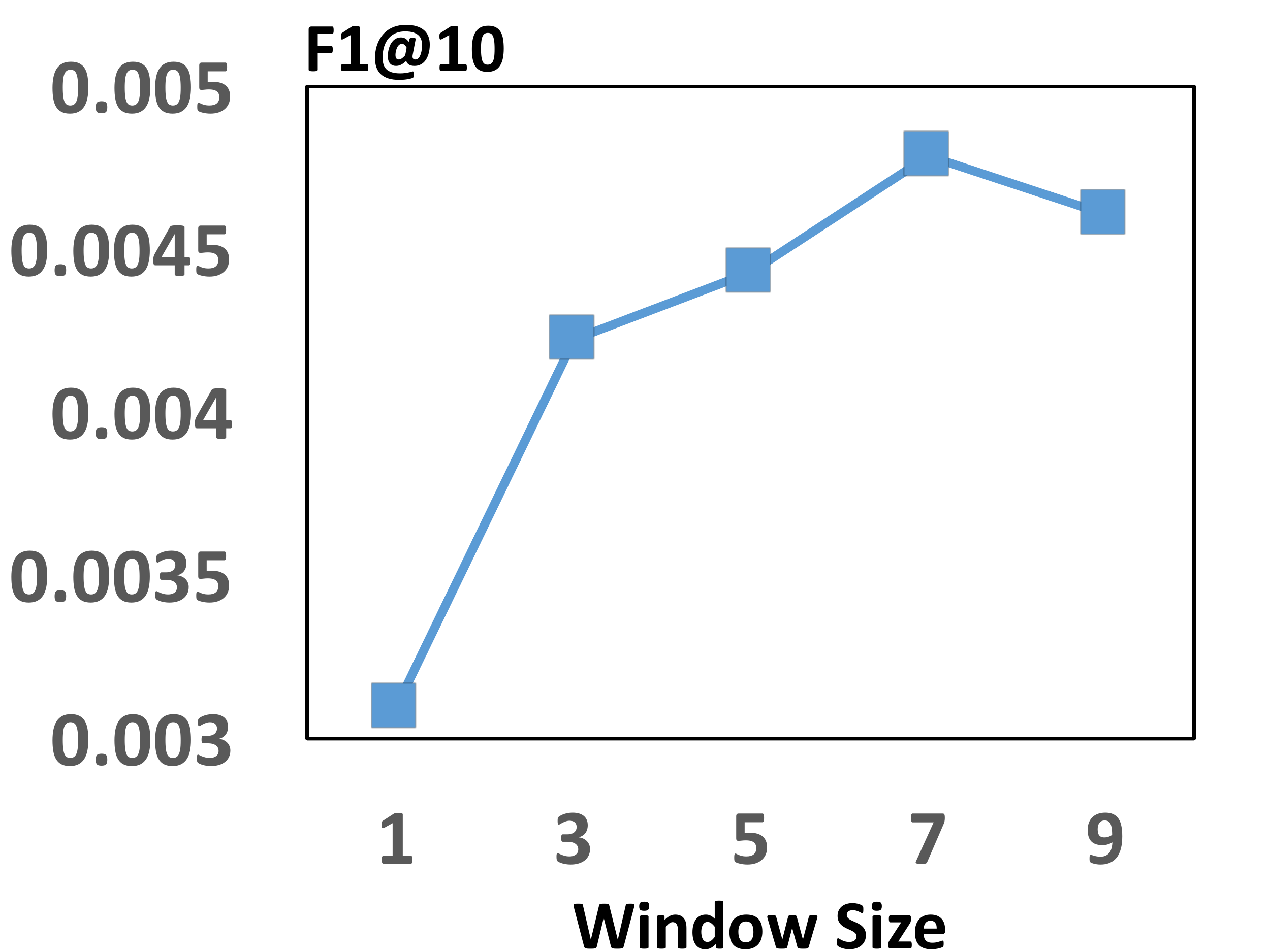}}
   %% label for entire figure
  \subfigure[Interactions Kept: $40\%$]{
    \label{40} %% label for second subfigure
    \includegraphics[width=1.3in]{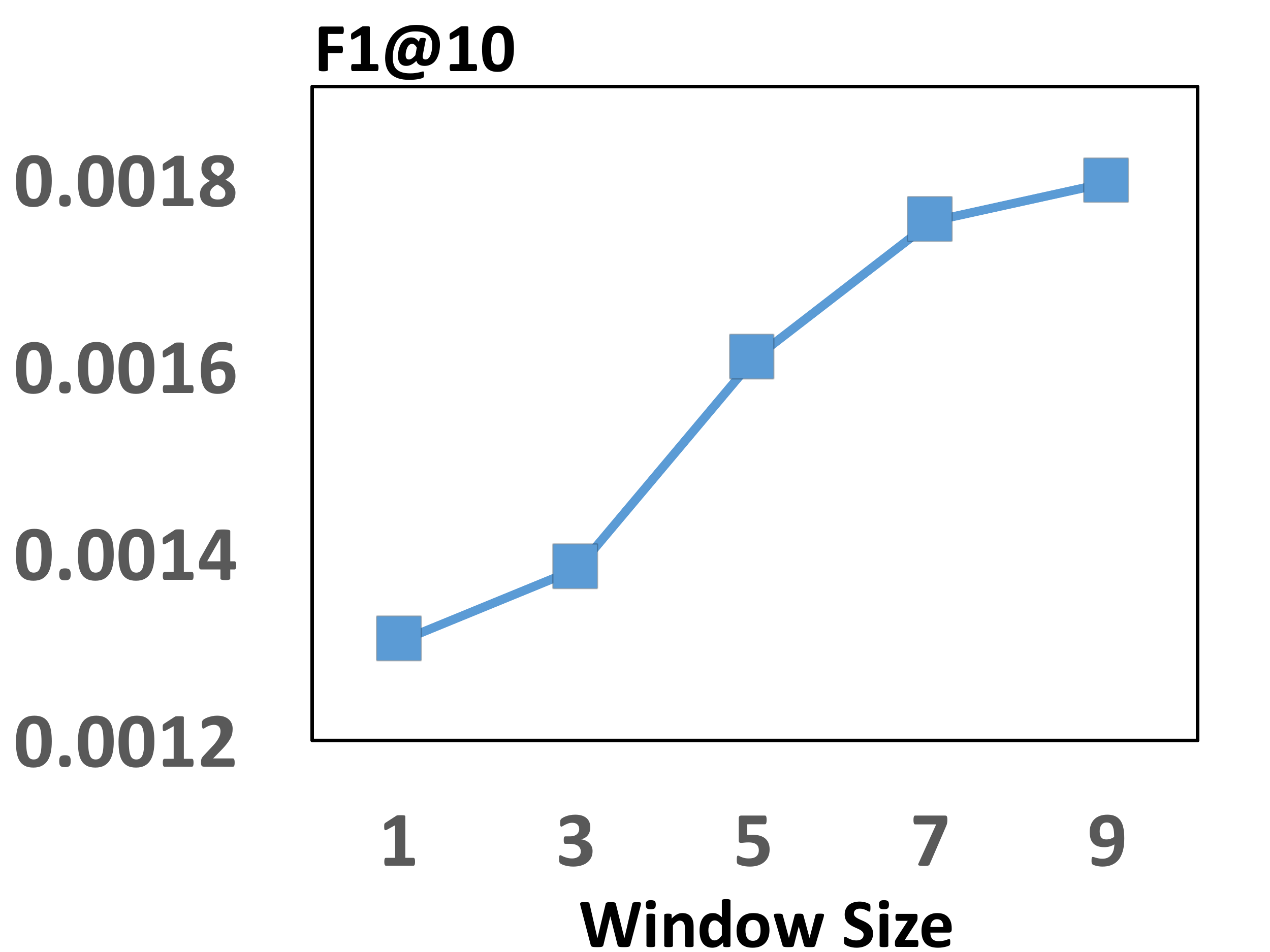}}
    %% label for entire figure
   \subfigure[Interactions Kept: $20\%$]{
    \label{20} %% label for second subfigure
    \includegraphics[width=1.3in]{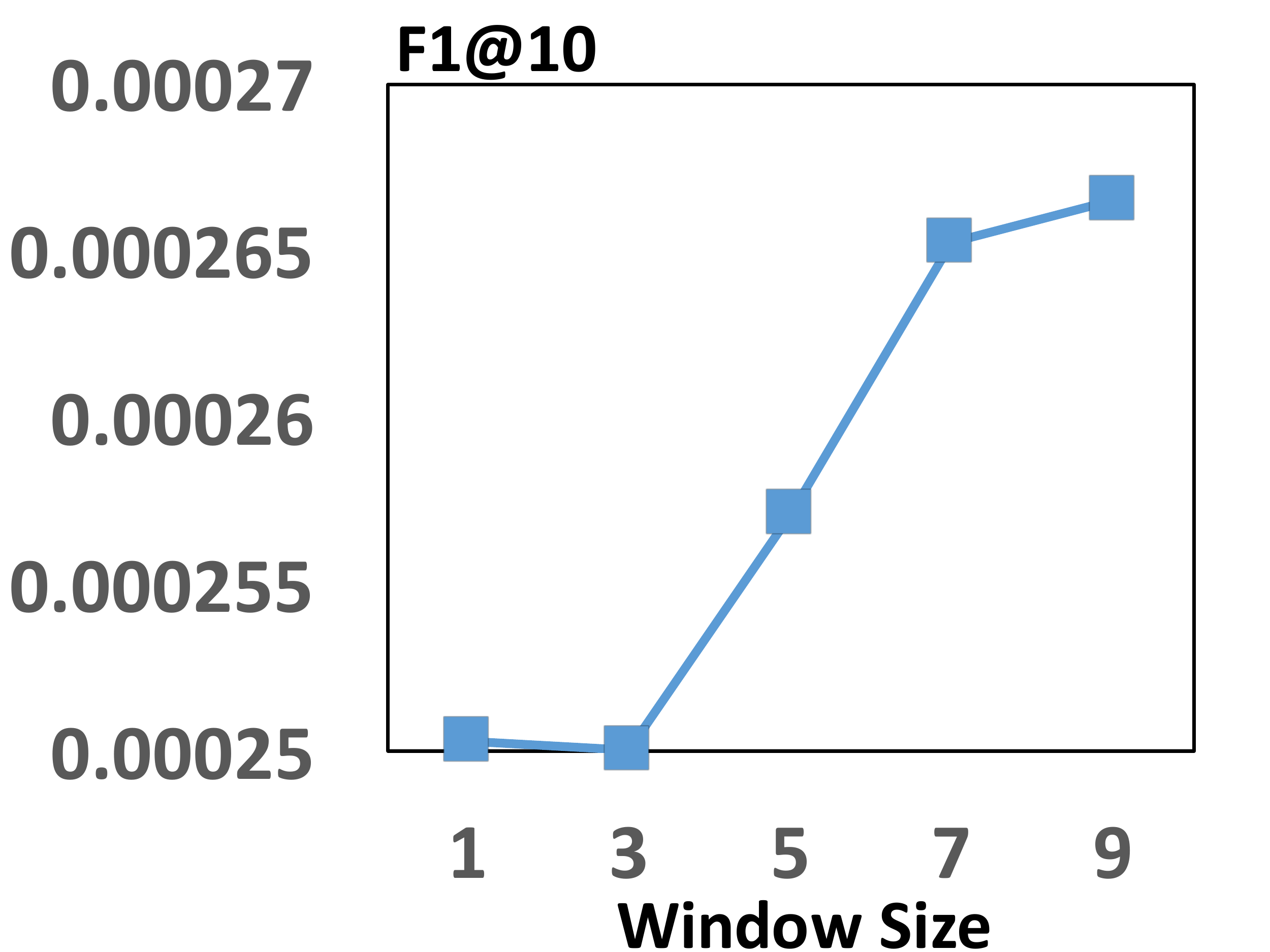}}
  \caption{Impact of the window size on Amazon Toys and Games dataset and its four sparser versions.}
  \label{Impact of the window size on different sparse datasets} %% label for entire figure
\end{figure*}

\subsection{Robustness of \textit{PsiRec-PMI} on Very Sparse Datasets}
\label{Robustness of Our Approach on Very Sparse Dataset}
\begin{figure}[htbp]
\centering
\includegraphics[width=2.1 in]{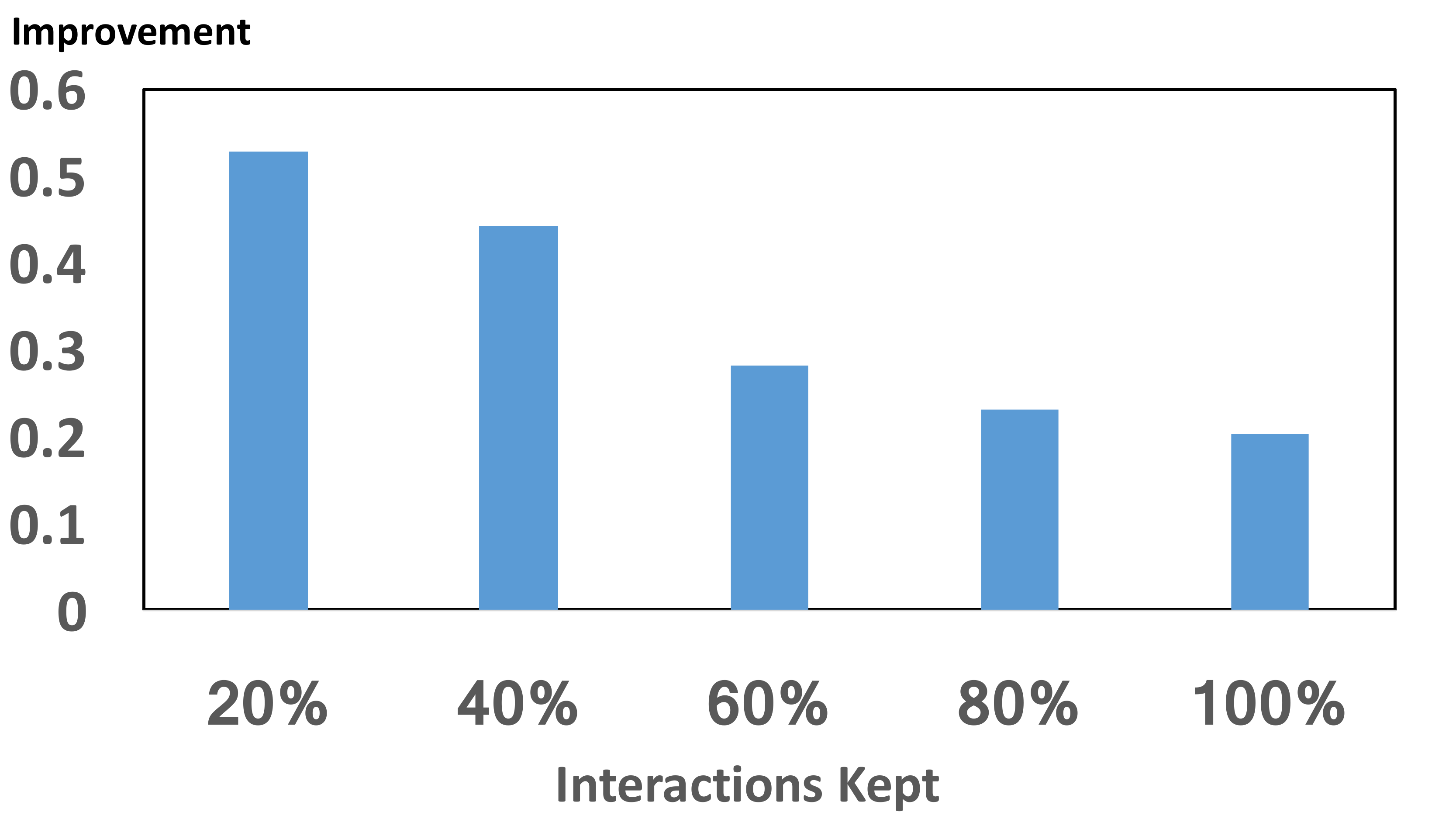}
\caption{Improvements (F1@10) of PsiRec upon CDAE on Amazon Toys and Games and its four sparser versions.}\label{Improvements upon CDAE on the Sparse Datasets}
\end{figure}

We further evaluate the effectiveness of \textit{PsiRec-PMI} as the degree of data sparsity increases. Keeping the same set of users and items, four sparser versions of the Amazon Toys and Games dataset are generated by randomly removing 20\%, 40\%, 60\% and 80\% purchases for each user. Particularly, Figure \ref{Improvements upon CDAE on the Sparse Datasets} shows that the sparser the datasets are, the larger improvements in terms of F1@10 are achieved upon the strongest baseline CDAE. This demonstrates the promising ability of PsiRec to alleviate the data sparsity problem. %We attribute the success on sparse datasets to the pseudo-implicit feedback matrix $\textbf{S}$ generated by PsiRec-PMI which is denser than $\textbf{R}$.

%We evaluate \textit{PsiRec-PMI} and the baselines on the four sparser datasets and report the results in terms of F1@10 in Figure \ref{Robustness on sparse datasets}. %The statistics of the four sparser versions and the original dataset is presented in Table \ref{toy sparser}.
%
%The observations from Figure \ref{Robustness on sparse datasets} are two-fold. The first observation is that when the dataset sparsity increases, there is a significant drop in performance for all methods. This indicates that sparse datasets are challenging for all methods since there are fewer user-item interactions available to learn user preferences. The second observation is that \textit{PsiRec-PMI} is always the best even on these sparser datasets. 

% Table generated by Excel2LaTeX from sheet 'toy sparser'
%\begin{table}[htbp]
%  \centering
%	\setlength{\tabcolsep}{3.0pt}
%  \caption{Sparser Versions of Amazon Toys and Games DataSet}
%    \begin{tabular}{|c|ccccc|}
%	\toprule
%    Interactions Kept  & 20\% & 40\%	 & 60\%	& 80\% & 100\%\\
%	\midrule
%    Interactions\# & 38,941 & 68,334 & 101,096 & 133,859 & 167,597\\
%    Density   & 0.017\% & 0.030\% & 0.044\% & 0.057\% & 0.072\%\\
%	\bottomrule
%    \end{tabular}%
%  \label{toy sparser}%
%\end{table}%

%\begin{figure}[htbp]
%\centering
%\includegraphics[width=3.2 in]{image/toy_sparser.pdf}
%\caption{Results (F1@10) of PsiRec and baseline methods on Amazon Toys and Games dataset and its four sparser versions.}\label{Robustness on sparse datasets}
%\end{figure}

\subsection{Impact of Window Size}
\label{Impact of Window Size}

\subsubsection{Effectiveness of Transitive User-Item Relationship}
\label{Effectiveness of Transitive User-Item Relationship}
In the user-item bipartite graph, indirectly connected user-item pairs will be sampled by our random walks only if $\sigma>=3$. Figure \ref{Impact of the window size on different sparse datasets} shows the performance of \textit{PsiRec-PMI} as we vary this window size on the Amazon Toys and Games dataset and its four sparser versions. We observe that the optimal result in terms of F1@10 is always achieved when $\sigma>=3$, with a considerable improvement upon $\sigma=1$ on all datasets. This indicates that, on sparse datasets, indirectly connected user-item pairs can significantly enhance matrix factorization methods for top-K recommendation. %Therefore, we suggest incorporating transitive user-item relationships to estimate the actual user preference on extremely sparse datasets.

%\subsubsection{Impact of Window Size on Density of $\textbf{S}$}
%The window size $\sigma$ is a key factor for sampling the user-item pairs. On the one hand, the larger the window size, the more indirect user-item pairs will be sampled. Consequently, the density of the preference matrix $\textbf{S}$ will increase. An example of this on Amazon Toys and Games dataset is presented in Table \ref{window size density}. Specifically, the larger the window size, the more indirect user-item pairs will be sampled to enrich $\textbf{\textbf{S}}$. %On the other hand, the larger the window size, the indirect user-item pair which has a longer distance in the user item bipartite graph will be sampled, which has been shown in Section \ref{Sampling of the User-item Pairs}. 

%\begin{table}[htbp]
%  \centering
%  \caption{Impact of the Window Size on the Density of $S$ on Amazon Toys and Games Dataset}
%    \begin{tabular}{|c|ccccc|}
%	\toprule
%        Window Size  & 1     & 3     & 5     & 7     & 9 \\
%	\hline
%    Density(\%) & 0.058 & 3.696 & 8.323 & 12.805 & 16.984 \\
%	\bottomrule
%    \end{tabular}%
%  \label{window size density}%
%\end{table}%

\subsubsection{Optimal Window Size on Sparse Datasets}
The impact of the window size on \textit{PsiRec-PMI} on Amazon Toys dataset and its four sparser versions is shown in Figure \ref{Impact of the window size on different sparse datasets}. There are two observations: (1) The performance is sensitive to the window size on all datasets; and (2) The sparser the dataset is, the larger the optimal window size is. Specifically, the optimal window size for datasets keeping percentage: 100\%, 80\%, 60\%, 40\% and 20\% are 3, 5, 7, 9, 9 respectively. This suggests that the sparser the dataset, a larger window size is needed to sample more and longer transitive user-item relationships to enrich $\textbf{S}$ so that the user preferences can be better estimated.

\section{Conclusions and Future Work}
\label{Conclusion and Future Work}
We propose PsiRec, a user preference propagation recommender designed to alleviate the data sparsity problem in top-K recommendation for implicit feedback datasets. Extensive experiments show that the proposed matrix factorization method significantly outperforms several state-of-the-art neural methods, and that introducing indirect relationships leads to a large boost in top-K recommendation performance. In the future, we aim to improve PsiRec by exploring pairwise learners and the impact of incorporating side information like  review text and temporal signals.% into PsiRec. %The intuition of PsiRec is to extend the original sparse dataset by incorporating indirect, transitive user-item relationships as pseudo-implicit feedback. We first view the interactions as a bipartite graph and use random walks to extract indirect user-item pairs from the graph. After that, pseudo-implicit feedback is generated from the indirect user-item pairs. To measure the confidence of pseudo-implicit feedback and motivated by the Skip-gram model, we first propose an embedding model that operates on the individual user-item pairs. We then show that this model is implicitly factorizing an PMI matrix between users and items. Finally, a latent factor model is used to estimate the user preference from the pseudo-implicit feedback.

\bibliographystyle{IEEEtran}
\bibliography{main}
\end{document}